\documentstyle[preprint,aps, epsf]{revtex}

\begin{document}

\title{\bf Scaling Behaviour of Conformal Fields in Curved Three-dimensional Space}

\author{George Tsoupros \\
       {\em School of Mathematics and Statistics}\\
       {\em The University of Sydney}\\
       {\em NSW 2006}\\
       {\em Australia}
\footnote{present e-mail address: georget@maths.usyd.edu.au}}

\maketitle

\begin{abstract}

The limitations of three-dimensional semi-classical gravity are explored in the context of a conformally invariant
 theory for a self-interacting scalar field. The analysis of the theory's scaling behaviour reveals that 
 scalar-loop effects contribute to the conformal anomaly only at advanced orders allowing for a
range of relevant energy scales which extend to those comparable to the non-perturbative scale of Planck mass.

\end{abstract}

\newpage

%{\bf I. Renormalisation in the Effective Action}\\

The validity of a semiclassical approach which differentiates between the significance of the
gravitational and matter degrees of freedom in the context of second quantisation is an important
issue in perturbative quantum gravity. The conventional approach to semiclassical gravity
specified by a classical metric coupled to the quantised matter fields with the expectation value of
 the relevant stress tensor acting as a source to the Einstein equations is contingent upon the
 validity of the semiclassical expansion of the theory's generating functional. That loop expansion,
 in turn, depends on the energy scale of the problem \cite{Birell} which can, itself, be determined
 from the character of the effective theory. The latter is specified by the quantum
  state of the system and the location of maxima of the corresponding effective action. In effect, information about the energy scales relevant to the validity of the semiclassical expansion is contained in the renormalisation group behaviour of the effective theory. This letter
addresses this issue in the case of a conformally invariant, self-interacting  scalar field theory in
 curved three-dimensional spacetime. The perturbative renormalisation of this theory has been
  considered in a space of constant curvature \cite{McKeon Tsoupros} which allows for the
 exploitation of the fact that in three dimensions all divergences induced in the gravitational action
 by scalar vacuum effects can only be proportional to the scalar curvature and cosmological
 constant $ \Lambda$. The following analysis advances the renormalisation of the theory in the effective action and 
tackles the issue of the validity of the underlying semiclassical approximation by exploring the implications of 
 the renormalisation group behaviour of the effective theory. It is shown that the effective theory is characterised by 
the absence of any contribution to the conformal anomaly up to third order in the loop-counting parameter $ \hbar$. Scalar vacuum effects generate conformally non-invariant counterterms at orders no smaller than the fourth in the loop expansion and they, thereby, determine the range of energy scales relevant to the validity of the semi-classical expansion.   
As the curvature of space-time is not
 expected to have an impact on the cut-off scale behaviour of the theory manifest in
 the leading divergences of the Green functions, its scaling behaviour toward the high-energy
 limit will be independent of the background geometry. 

The spherical formulation of the dynamical behaviour of a self-interacting scalar field $\phi$
  admitting a classical action which remains
 invariant under the conformal transformation
 \cite{Wald}
 
\begin{equation}
g_{\mu\nu} \rightarrow \Omega^{2}(x)g_{\mu\nu},~~~
\phi \rightarrow \Omega^{1-\frac{1}{2}n}\phi \equiv \Phi
\end{equation}
in a general n-dimensional Lorentzian spacetime,
is attained on a Riemmanian manifold by specifying the background geometry to be that of
 positive constant curvature $ r$ embedded in a (n+1)-dimensional space. With the embedded
sphere's surface element being $ d{\sigma}=r^nd{\Omega}_{n+1} $ and with the conformal
 scalar field $ \Phi$, specified on it, admitting the only possible self-interaction for a renormalisable theory 
at $ n=3$, the scalar action
 reduces to \cite{McKeon Tsoupros}

\begin{equation}
I[{\Phi}] = \int{d\sigma}[\frac{1}{2} \frac{1}{2r^2}
\Phi(L^2- \frac{1}{2}n(n-2) )\Phi - \frac{\lambda}{6!}\Phi^6]
\end{equation}
The differential operator $ L_{ab} $ is the generator of rotations on the surface of the
n-dimensional sphere. The scalar curvature $R $ of the latter relates to the radius $r $ through

\begin{equation}
R = \frac{n(n-1)}{r^2}
\end{equation}
This Riemannian manifold is characterised by spherical (n-1)-dimensional sections of constant
 Euclidean time $ \tau$. The gravitational component of the action functional for the Euclidean sphere's 
relevant segment $ M$ characterised by $ g_{\mu\nu}$ and bounded by an (n-1)-dimensional hypersurface $ \partial M$ at  
 $ n=3$ is 

\begin{equation}
I_g =   
2\int_{{\partial}M}{d^2}x \sqrt{h}K
+ \frac{1}{\kappa^2} \int_{M}{d^3}x \sqrt{g}(R -2\Lambda) 
\end{equation}
where an irrelevant parity violating Chern-Simons term has been omitted \cite{McKeon Tsoupros}.
With $ h_{ij}$ being the metric of the two-dimensional hypersurface of constant Euclidean time 
$ \tau$, the surface term involves the trace $ K$ of the extrinsic curvature $ K_{ij}$ of the
 boundary two-sphere and the coefficient $ \kappa^{-2}$ has dimensions of mass. The assumption of spatial homogeneity for the scalar field $ \Phi(x)=\Phi(\tau)$
 forces the additional surface term, effected on $ \partial M$ from the scalar action, to vanish.

The generating functional for this theory in a space of Euclidean signature is  

\begin{equation}
Z[J=0] = 
\int{D{\Phi}}~e^{-\frac{1}{\hbar}I[g_{\mu\nu}, \Phi]}
\end{equation}
with the Euclidean action functional at the dimensional limit $ n = 3$ being the additive result
of $ I[\Phi]$ and $ I_g$. The semi-classical ( $\hbar \rightarrow 0$ ) expansion of $ I[g_{\mu\nu}, \Phi]$ is 
based on the classical extremal $ \Phi^{(cl)}$:

$$
I[g_{\mu\nu}, \Phi] = I[g_{\mu\nu}, \Phi^{(cl)}] + \frac{1}{2!}<(\Phi - \Phi^{(cl)})I^{(2)}(\Phi - \Phi^{(cl)})>_{xy} + ...~~~~ ;$$

\begin{equation}
\frac{{\delta}I[\Phi^{(cl)}]}{{\delta}{\Phi}}_{|M} = 0,  \Phi^{(cl)}(\tau)_{|M} = reg.,   
\Phi^{(cl)}(\tau)_{|{\partial}M} =  \Phi_{|{\partial}M}
\end{equation}
where the brackets $ <>$ denote invariant integration over spacetime variables and
the successive derivatives $ I^{(m)}$ (m = 2, 3, ...) of the action functional are evaluated 
at $ (g_{\mu\nu}, \Phi^{(cl)})$. Substituting in (5), it results in  

\begin{equation}
Z[J=0] =
e^{-\frac{1}{\hbar} \int_{{\partial}M}{d^2}x\sqrt{h}K } 
e^{- \frac{1}{\hbar}\Gamma[\Phi^{(cl)}]}
\end{equation}

with

$$
\Gamma[\Phi^{(cl)}] =  \int_M{d\sigma}\frac{1}{\kappa^{\rm 2}_{0}}[R - 2\Lambda_{0}] + \int_M{d\sigma}[\frac{1}{2} \frac{1}{2r^2}
\Phi_{0}(L^2- \frac{1}{2}n(n-2) )\Phi_{0} - \frac{\lambda_{0}}{6!}\Phi_{0}^6] $$
\begin{equation}
+ \sum_{n=1}^{\infty} \hbar^n L^{(n)}(\Phi^{(cl)}) 
\end{equation}
The exponentiated expression $ \Gamma[\Phi^{(cl)}]$ whose perturbative 
expansion in terms of $ \hbar$ appears in (8) is necessarilly
the renormalised effective action of the present semi-classical theory. Being the
 generating functional for proper vertices, it can be calculated perturbatively by summing over
 vacuum diagrams plus counterterms to any specific order \cite{Coleman}. Such diagramatic summation to all
orders amounts to expressing the effective action in terms of the standard ``classical field''. The (unrenormalised) contributions $ L^{(n)}(\Phi^{(cl)})$ to each loop-order relate to the derivatives $ I^{(m)}$ 
through functional integrations stemming from (5). (For instance, the one-loop effective action is expressed, through
Gaussian integration,  by the 
usual result $ L^{(1)}(\Phi^{(cl)}) = \frac{1}{2}trln[I^{(2)}(\Phi^{(cl)})]$ ).    
The bare quantities $ (\Lambda_{0}, \Phi_{0})$ which appear in the action contain the counterterms which will cancel
 identically against the divergences contained in $ L^{(n)}$ successively in $ \hbar$. The perturbative evaluation of the effective action will be predicated, in what follows, on the use of dimensional regularisation \cite{t'Hooft} which 
manifests the divergences as poles at the
 dimensional limit of $ n=3$. With minimal subtraction of the divergences and with $\epsilon=3-n$ the relevant
 't Hooft expansions for the bare parameters are

\begin{eqnarray}
\kappa^{\rm -2}_{0}=\mu^{3-n}[ \kappa^{\rm -2}+\sum_{k=1}^{\infty}\sum_{i=1}^{\infty}
\frac{\beta_{ki}\lambda^i}{\epsilon^k}] ;~~
\Lambda_0=\mu^{3-n}[\Lambda+\sum_{k=1}^{\infty}\sum_{i=1}^{\infty}
\frac{\alpha_{ki}\lambda^i}{\epsilon^k}] ; 
\nonumber \\
\Phi_0=Z^{1/2}\Phi^{(cl)} ; ~~
Z=1+\sum_{k=1}^{\infty}\sum_{i=1}^{\infty}
\frac{c_{ki}\lambda^i}{\epsilon^k} ;~~
\lambda_0=\mu^{3-n}[\lambda+\sum_{k=1}^{\infty}\sum_{i=1}^{\infty}
\frac{a_{ki}\lambda^i}{\epsilon^k}]
\end{eqnarray}
 The three-dimensional spherical formulation of this
 semi-classical theory allows for the perturbative evaluation of any Green function as an exact
 function of the positive constant $ R$ \cite{McKeon Tsoupros}.
The pole structures contained in $ L^{(n)}(\Phi^{(cl)})$ are essentially the sum total of contributions,
 arising perturbatively in $ \hbar$, from all connected diagrams. Demanding that $ 
 \Gamma[\Phi^{(cl)}]$ be finite as $ n \rightarrow 3$ fixes the pole 
 structures and residues of the counterterms contained perturbatively in (8) to the values 
 cited in \cite{McKeon Tsoupros}.  Supplementing these results with their associated finite parts stemming from  
all irreducible diagrams 
 up to order $ \lambda^3$ 
 the radiative contributions to the bare action translate successively to 
the following contributions in the loop expansion of the effective action

$$ \hspace{3in} L^{(1)}=0 \hspace{2in} ~~~~~~~~~~(10a)$$
which is a direct consequence of the vanishing effect dimensional regularisation 
has on the massless tadpoles
 \cite{Drummond},
$$ \hspace{2.5in} L^{(2)}=\frac{5\lambda^2}{96\pi^2}\frac{1}{\epsilon}+\lambda^2 \frac{5}{96\pi^2}(2ln{\mu} - \gamma_E)
 \hspace{0.5in} ~~~~~~~~(10b)$$ 
$$ L^{(4)}=\frac{\lambda_0^2}{120}\left[\frac{1}{\pi^4 . 3 . 2^{11}}\right][(\frac{L^2-\frac{3}{2}}{2r^2}) \frac{1}{\epsilon}+ \frac{3}{2r^2}(- \frac{5}{2}+$$ 
$$\hspace{1.5in}
\frac{L^2}{9}+(\frac{L^2}{3}-\frac{1}{2})(-3+2ln{\pi}+4ln(2r)+4ln4+4\gamma_E))+0(\epsilon)]
\hspace{0.5in} ~~~~~~(10c)$$
with its finite part at $\epsilon \rightarrow 0$ being
$$
\lambda^3\frac{1}{\pi^6}\frac{1}{2^{19} .~ 3^3}\frac{1}{2r^2}[(L^2(2ln{\mu}-\frac{8}{3}+2ln{\pi}+4ln(2r)+4ln4+4\gamma_E)-$$
$$3ln{\mu}-3ln{\pi}-6ln(2r)-6ln4-6\gamma_E-3]+F.T.$$
$$ \hspace{2in}
L^{(5)} = \tilde{L}^{(5)} + \frac{\lambda^3}{\pi^6 r^3} ~\left[\frac{1}{96 . 144 . 3 . 2^{12}}\right]
\hspace{1.5in} ~~~~(10d) $$
with $ \tilde{L}^{(5)}$ stemming from the contributions of all four-point diagrams relevant to $O(\hbar^5)$ and 
satisfying $ \tilde{L}^{(5)}+L^{(3)}=0$ as a direct consequence of the absence of any divergence in the 
four-point function to $ O(\lambda^3)$ \cite{McKeon Tsoupros},
$$ 
L^{(6)} = -\frac{\lambda^3}{72} \frac{1}{3 . 4^7. \pi^6 . \epsilon^2}  \left[1+\epsilon[6ln(2r)+3ln{\pi}+
6(ln4+\gamma_E)-5-3\gamma_E]\right] . $$
$$\hspace{2in}
\left[(-\frac{1}{4}+\frac{L^2}{6})+\epsilon(-\frac{9}{4}+
\frac{2L^2}{9})\right]
\hspace{1.8in} ~~(10e) $$
with its finite part at $\epsilon \rightarrow 0$ being
$$
-\frac{1}{2r^2} \frac{\lambda^3}{72} \frac{1}{4^7 . \pi^6}[L^2(\frac{2}{9}+\frac{1}{6}[6ln(2r)+3ln{\pi}+6ln4+3\gamma_E-5]+\frac{1}{2}ln{\mu})ln{\mu}+
$$
$$
(-\frac{3}{4}ln{\mu}-\frac{3}{2}ln(2r)-\frac{3}{4}ln{\pi}-\frac{3}{2}ln4-\frac{3}{2}\gamma_E-5)ln{\mu}
-\frac{9}{4}ln{\mu})+\tilde{L}^{(6)}+F.T.
$$
with $ \tilde{L}^{(6)}$ effected by the finite contributions arising at $ O(\lambda^4)$. Finally, 
$$
\hspace{2in}
L^{(7)}=\lambda^3\frac{1}{3 . 1296 . ~2^{15}}\frac{1}{\pi^6 . r^3}
\hspace{2in}~~~~~(10f)$$
In all these expressions the remaining finite terms (F.T.)
are $\mu$-independent. In addition, multiplication of each loop-contribution, arising from any connected diagram, 
by that power of the renormalised scalar field $ \Phi_R$ which coincides with that diagram's number of external legs,
is implied in the effective action. As is evident from (8), renormalisation in the effective action necessitates the expression of $ \Phi_R$ in terms of $ \Phi^{(cl)}$.  
It is understood that, in a general renormalisation scheme, all terms in the loop 
expansion of $ \Phi_R$ in terms of the semi-classical configuration $ \Phi^{(cl)}$ are determined
through finite redefinitions of the latter in perturbation theory, a situation inherent in the expression for 
$ \Phi_0$ in (9)(with $ \Phi_R$ being the product of the sum-total of all finite terms in 
$ Z^{\frac{1}{2}}$ times $ \Phi^{(cl)}$).
In minimal
 subtraction all counterterms are characterised by the absence of finite parts, so that $ \Phi_R$ 
coincides identically with $ \Phi^{(cl)}$. 
In effect, each loop
 contribution in (10) is multiplied by that power of $ \Phi^{(cl)}$ which is determined by the number of
 external legs for the corresponding diagrams. In accordance with the results in \cite{McKeon Tsoupros} the 
divergent contributions to the two-point function are absorbed in a wavefunction renormalisation 
of the scalar field, generating $ \Phi_0$ to orders $ \hbar^4$ and $ \hbar^6$ respectively. This follows 
the self-coupling renormalisation, which stems from the divergent contribution to the six-point function. 
The counterterms thus generated through (9), cancel identically against the corresponding pole structures in the 
loop expansion (8) resulting in finite contributions to the semi-classical exponentiated 
effective action, specified perturbatively by the finite parts in (10). As stated, to order $ \lambda^3$ 
no divergence arises in the four-point function and no conformally non-invariant counterterm 
proportional to $ R\phi^2$ is induced by vacuum effects. The contributions to the zero-point 
function result in finite redefinitions of the cosmological constant at orders $ \hbar^5$ 
and $ \hbar^7$  respectively.  
%Comparison with (8) shows that to order 
  %$ \hbar^5$ and $ \hbar^7$ there are finite contributions to the cosmological constant 
  %$  \Lambda$. 
 %In conformity with (8) and (13), (14) the effective cosmological constant is: 
%
%\addtocounter{equation}{1}%
%\begin{equation}
%\frac{1}{\kappa_{0}^{\bf 2}} \Lambda_{0} = \frac{1}{\kappa^{2}} \Lambda +  \lambda^3 %\frac{1}{r^3} \frac{1}{\pi^6 . 3 . 2^{12}}
%\left[ \hbar^5 \frac{1}{96 . 144} + \hbar^7 \frac{1}{2^3 . 1296} \right]
%\end{equation}
%
%Finally, wavefunction renormalisation for the conformal field, expressed in (13) and (18), stems
 %from the additive result of diagrams (e) and (g) to order $ \hbar^4$ and $ \hbar^6$ %respectively.  After the relevant cancellations against the pole structures in (14c) and (14e) the %contribution to the kinetic sector of the effective action in (12

%{\bigskip}
%{\bf II. Scaling and High Energy Behaviour}\\

As discussed in the introduction, the validity of the semi-classical approach which retains the background geometry fixed at a classical configuration essentially depends on the energy scale of the problem and can be determined by the behaviour of the semi-classical effective action of the theory. For an n-dimensional local field theory the classical action has the asymptotic scaling invariance 

\addtocounter{equation}{1}%
\begin{equation}
I[\tilde{{\bf q}}] = I[{\bf q}],~~~ \Omega(x) \rightarrow 0
\end{equation}
under the conformal transformations (1) on $ {\bf q} = (g_{\mu \nu}(x), \phi(x))$. On the contrary, at the dimensional limit $ n$
covariantly regulated quantum corrections result in the anomalous
scaling behaviour for the one-loop effective action \cite{DeW}
%
%\begin{equation}
%\int{d^n}x\left[2\tilde{g}_{\mu \nu}\frac{\delta}{\delta{\tilde{g}_{\mu \nu}}} + 
%(1-\frac{n}{2})\tilde{\phi}\frac{\delta}{\delta{\tilde{\phi}}}\right] ~\frac{1}{2}TrlnF[\tilde{{\bf q}}]
%= -\frac{1}{16\pi^2}A_2[{\bf q}],~~~\Omega(x) \rightarrow 0
%\end{equation}
%representing the conformal anomaly of the quantum theory and defined by the
 %coefficient $ A_{2}$ of the Schwinger-De Witt proper-time expansion for the functional trace of
 %the heat kernel of the operator $ F[\tilde{{\bf q}}]$ \cite{DeW}. Thus, it follows from (23) that in %the limit of small distances the scaling behaviour of  the one-loop effective action

\begin{equation}
\Gamma_{1-L}[\tilde{{\bf q}}] = 
-\frac{1}{16\pi^2}A_2[{\bf q}]ln{\Omega}\Gamma_{1-L}[{\bf q}],~~~\Omega(x) \rightarrow 0
\end{equation}
This scaling behaviour is, in effect, determined by the coefficient $ A_2$ associated with the heat kernel of the conformal field $ \Phi$.  The conformal anomaly is expressed by the volume density of $ A_2$. The validity
 of the corresponding semi-classical expansion of the functional integral is determined by
 the extent to which the value of $ A_2[{\bf q}]$ in the exponentiated effective action suppresses
 the contribution of higher orders in quantum gravity, stemming from the high-energy scales as $
 \Omega(x) \rightarrow 0$, to result in a convergent expression for the path integral. That perturbative
 expansion ceases to be reliable at distance-scales at which the functional integral $ Z[J=0]$ does not admit such an expression. Consequently, the distance-limit beyond which the expansion breaks down is 
 determined by the anomalous scaling behaviour of the one-loop effective action and signals the
 scale at which higher-order gravitational effects become important. The anomalous scaling is
 itself controlled, perturbatively, by the anomalous dimension manifest at that high-energy scale
 in the proper functions of the theory and resulting in the conformal anomaly, as is evident in (12).
This analysis applies directly to the present three-dimensional case. The effective theory is described by the 
renormalised effective action derived up to order $ \hbar^7$. Its behaviour under the conformal rescalings (1)
toward the asymptotic limit of $ \Omega \rightarrow 0$ contains information about the energy-scales relevant to the generation of conformally non-invariant quantum effects and, thereby, relevant to the limitations of the semi-classical expansion.      
The renormalisation
  program is indicative of the
 background geometry's irrelevance to the scaling behaviour of the theory toward the high-energy limit. For
 instance, the term in (9) which expresses the two-loop
 contribution to $ \lambda_0$
%
%\begin{equation}
%\frac{a_{11}\lambda}{\epsilon} = \frac{5\lambda^2}{96\pi^2}\frac{1}{\epsilon}
%\end{equation}
 is identical to that effected in the flat space-time formulation of the $ \phi^6$ scalar theory
 \cite{McKeon Tsoupros}, \cite{MK T}. This result is in conformity with intuitive expectation.
The leading divergence inherent in a diagram, is a localised effect arising from the coincidence of all 
integration points and is,
 consequently, not expected to be affected by the radius of the background geometry. The ensuing considerations 
have, for that matter, general validity in any space-time. 

Despite the stated absence of those conformally non-invariant counterterms which - to order $ \lambda^3$ - are associated 
with a divergence in the four-point function as well as with the $ R \phi^2$ sector, the potential for conformal breaking as a result of the anomalous scaling has to be explored independently in the expansion of the effective action up to 
order $ \hbar^7$. The latter is the highest loop-order associated with the diagrams relevant to order $ \lambda^3$. 
The result expressed by (10a) has the general significance of the complete absence 
of divergences at 
 one-loop level in odd dimensionalities. The one-loop effective action is finite and its 
 corresponding contribution to the conformal anomaly vanishes \cite{Birell}. The two-loop contribution to the 
renormalised effective action 
 is expressed by the finite part of $ [L^{(2)}][\Phi^{(cl)}]^6$ in (10b). 
However, the anomalous dimension is controlled 
 exponentially by the residues $ c_{ki}$ of the counterterms in (8) for field renormalisation \cite{Brown}. 
Since $ c_{ki}$ are, as stated, of loop-order no smaller than $ \hbar^4$, the contribution 
of the two-loop effective action to the
 conformal anomaly is vanishing. (This result is identical to that in \cite{Odintsov} obtained at this loop-order from the examination of the $\beta$-function of the theory.) A similar argument holds for the three-loop 
 contribution to the effective action. Consequently, conformal invariance persists at least up to third order in the
loop expansion as scalar-loop effects generate conformally non-invariant counterterms only at higher orders. These counterterms determine, in turn, the perturbative contributions to the anomalous scaling of the theory's 
Green functions and, thereby, to the conformal anomaly. The validity of the semi-classical expansion in the limit of small distances depends on the character of the anomalous scaling and on that of the conformal anomaly. The conformally non-invariant counterterms effectively determine
the range of energy scales whithin which the semi-classical approximation to quantum gravity is valid. 
 The semi-classical expansion of the theory's generating functional, in (7) and (8), ceases to be reliable at energy scales at which the anomalous scaling of the operators violates perturbatively the requirement of a power falloff in the effective action. The relevant distance-limit signals the scale at which 
higher-order gravitational effects become important. For instance, (at one-loop) such a situation is generic in (12). Depending on the value of the corresponding contribution to the anomalous scaling determined by $ A_2[{\bf q}]$, the effective action either suppresses the contribution of the over-Planckian energy scales or infinetely enhances it and serves as an applicability criterion for the semi-classical expansion toward the high-energy limit $ \Omega \rightarrow 0$. 

In light of these considerations, the renormalisation group behaviour of the present theory calls the 
 semi-classical approximation into question at distance-scales at which the scalar-loop effects, 
responsible for the generation of conformally non-invariant counterterms, become significant.  
The fact that the validity of the expansion persists to high orders, suggests that the corresponding distance-scales are
 comparable to those of the $ \Omega \rightarrow 0$ limit and that, consequently, non-trivial 
 quantum gravitational effects enter at scales corresponding to hypersurfaces of vanishing two-geometry. Again, this 
result is independent of the background geometry. In the scalar case it could be  
conjectured, for that matter, that three-dimensional semi-classical gravity remains relevant at energy-scales 
comparable to the Planck mass.

\newpage

{\bf  Acknowledgements}
 
 I would like to acknowledge the contribution of critical theorist Dr. Ruth E. 
 Giensberg, professional contrarian and personal friend.

\end{document}